\documentclass[12pt,preprint]{aastex}

\newcommand{\be}{\begin{equation}}
\newcommand{\ee}{\end{equation}}

\begin{document}
\title{An Improved uvby-Metallicity Calibration for Metal-Rich Stars} 

\bigskip
\author{Sarah Martell, Gregory Laughlin} 

\bigskip
\affil{UCO/Lick Observatory, Department of Astronomy and Astrophysics, UCSC, Santa Cruz, CA 95064}

\begin{abstract} 
We present an improved uvby-metallicity relation calibrated for 
F,G, and early K dwarfs,
and an analogous uvby-$T_{eff}$ relation, both derived using a
Levenberg-Marquardt minimization scheme. Our calibrations are based on 1533 
stars which appear in both the Cayrel de Strobel (2001) metallicity 
compilation, and in the Hauck-Mermilliod (1998) catalog of uvby photometry.
We also examine the speculative possibility
of using uvby photometry to produce a uvby-planeticity calibration.
We conclude that while there is likely no strong photometric indicator 
of the presence or absence of short-period planets, the possibility of a
spectroscopic indicator of planeticity is well worth examining.

\end{abstract}


\section{Introduction} 
The number of known extrasolar planets is approaching 100, and more planets
are being discovered every month.\footnote {For the latest catalog of planets,
see http://www.exoplanets.org}
A remarkable empirical correlation has emerged from the aggregate of planets:
The parent stars of the detected extrasolar planets appear to be significantly
metal-rich in comparison to field FGK dwarfs in the solar neighborhood. 
This correlation was first extensively discussed by Gonzalez (1997),
and has been further studied by a number of authors; a partial list might
include Gimenez (2000), Laughlin (2000), Santos, Israelian \& Mayor (2001),
Murray \& Chaboyer (2001), and Reid (2002).  These studies all agree that
the fraction of stars containing readily detectable extrasolar planets
(e.g. having $K>15$  m/s, and $P<5$ yr) increases substantially
with increasing stellar metallicity. A search strategy geared toward the
most metal-rich stars should thus detect short-period planets at a 
far greater rate than a conventional volume-limited survey.

One might argue that a metallicity-based planet search strategy will
result in a haphazard accumulation of short-period planets,
while simultaneously
skewing the overall census of planets in the solar neighborhood. Indeed, it is
important to continue with volume-limited surveys in order to round out the 
unbiased statistical distribution of planetary systems. 
The discovery of additional
short-period planets, however, has important ramifications. For the following 
reasons, it is desirable to locate as many short-period planets as possible:
({\bf 1}) The presence of a short-period planet is an excellent 
indicator that additional planets are present in the system.
Currently, for example, 5 of 12 short-period planets 
observed at Lick Observatory show evidence for additional companions 
(Fischer et al 2001).
Recently, it has been realized that some multiple planet systems
such as GJ 876
allow for unambiguous determination of planet masses and orbital
parameters
(Laughlin \& Chambers 2001, Rivera \& Lissauer 2001).
({\bf 2}) The current
census of short-period extrasolar planets shows an interesting concentration of
objects in the period range between roughly 2.98 and 3.52 days 
(8 out of 20 planets with periods $P<20$ d). While
this pile-up is probably a consequence of disk-protoplanet tidal migration
(Lin, Bodenheimer \& Richardson 1996), the stopping mechanism is not at all
well understood. It is thus worthwhile to find more short-period planets to
better delineate the minimum planetary period. ({\bf 3}) 
Geometric arguments indicate
that 10\% of Hot Jupiter-type planets transit the face of the parent star.
A large increase in the rate of short-period planet detections 
thus translates into 
a significant increase in the number of transits discovered. 
Transiting planets permit
accurate mass and radius determinations, and are extremely 
important data points within
the overall theory of giant planets.

High-resolution spectroscopy provides the most precise method for flagging
the metal-rich stars which have a high probability of bearing a detectable planet.
Unfortunately, the number of field stars with spectroscopically determined metallicities
is relatively small.
The most recent compilation from the literature by Cayrel de Strobel, Soubiran,
\& Ralite (2001; hereafter CSR) contains 6534 metallicity determinations for 3356 stars.
The majority of single, metal-rich, main sequence and subgiant FGK stars in
the CSR compilation are already under radial velocity survelliance. 
A larger pool of high metallicity candidates is needed.

A study by Laughlin (2000) used the Hipparcos database (Perryman et al
1997), the Hauck-Mermilliod (1998) compilation of uvby photometry, and
the Schuster \& Nissen (1989; hereafter SN) uvby-[Fe/H] calibration to 
to identify 200 metal rich
stars from a parent population of roughly 10,000 K,G, 
and late F-type main sequence stars (having $7.8<V<10$)
in the immediate ($d<100{\rm pc}$) solar neighborhood. Nineteen
of these stars were added to the Keck Radial Velocity survey
in the Summer of 2000.
Three of these nineteen stars soon proved to be accompanied by planets
(Vogt et al 2001, Butler et al 2002).
Metallicity-based search strategies are 
therefore clearly optimizing the planet-detection efficiency.

The SN calibrations for F and G spectral types were based on 219 calibration
stars spanning a range of metallicities from
$-3.5 <  [{\rm Fe/H}] < 0.4$. This very useful calibration,
which has an overall standard
deviation of 0.16 dex per star, is wholly empirical, and has the 
advantage of using only the standard uvby indices, rather than the 
differential indices, $\delta m_{1}=m_{1}^{hyades}-m_{1}^{star}$, and
$\delta c_{1} = c_{1}^{hyades}-c_{1}^{star}$, which are used in the
earlier calibrations of Crawford (1975) and Olsen (1984). The SN
calibration has been widely used to investigate metallicity trends 
among solar neighborhood stars (e.g. Rocha-Pinto \& Maciel (1996), or
Haywood (2001). As recently emphasized by Twarog, Anthony-Twarog and Tanner
(2002), however, the SN calibration systematically underestimates
$[{\rm Fe/H}]$ by $\sim0.2$ dex for stars with spectroscopic metallicities
$[{\rm Fe/H}]>0.1$. The SN calibration is thus not optimally suited for
compiling an accurately ranked catalog of high metallicity stars. 

Our main goal in this letter is to make the uvby-metallicity calibration
more discerning and accurate among the highest metallicity
stars. We also use this forum to briefly introduce the
concept of a ``planeticity calibration'', in which stellar properties
are correlated directly with the detectable presence of an extrasolar
planet.
 
\section{New Photometric Calibrations} 

To produce our new calibration, we
we queried the CSR database for all stars which have (1)
Hipparcos parallaxes, (2) uvby measurements in the Hauck-Mermilliod (1998)
compilations, (3) ${\rm M_V} < 1.0$, and (4) lie within 100 pc of the Sun
(to reduce the problems posed by reddening). This subset of the CSR
catalog contains 1533 measurements of 664 stars. Using a Levenberg-Marquardt
scheme (see e.g. Press et al 1992), we then optimized the coefficients,
$A_{n}$, of the most general third-order polynomial in the Str\"omgren
indices $(b-y)$, $m_{1}$, and $c_{1}$ so as to best reproduce the
spectroscopically determined $[{\rm Fe/H}]$ values. Because the Levenberg-Marquardt
routine provides local convergence, it is prudent to start the
iterative sequence with different sets of initial guesses for ${A_{n}}_{i}$
to better insure that ${A_{n}}$, the set of final coefficients produced by the
routine, corresponds to the global minimum. We ran the Levenberg-Marquardt
procedure using (1) ${A_{n}}_{i}$ from the SN ``F star'' calibration 
(ignoring the logarithmic terms), (2) ${A_{n}}_{i}$ from the SN ``G star''
calibration, and (3) by setting all ${A_{n}}_{i}$ to unity. In each
case, the routine converged to an identical set of final coefficients, 
suggesting that a global minimum has been achieved. Our calibration is:

\begin{eqnarray}
[{\rm Fe/H}]_{phot} = &-&10.424602+31.059003(b-y)+42.184476m_1+15.351995c_1 \nonumber \\
&-&11.239435(b-y)^2-29.218135m_1^2-11.457610c_1^2-138.92376(b-y)m_1 \nonumber \\
&-&52.033290(b-y)c_1+11.259341m_1c_1-46.087731(b-y)^3+26.065099m_1^3 \nonumber \\
&-&1.1017830c_1^3+138.48588(b-y)^2m_1+39.012001(b-y)^2c_1 \nonumber \\
&+&23.225562m_1^2(b-y)-69.146876m_1^2c_1+20.456093c_1^2(b-y) \nonumber \\
&-&3.3302478c_1^2m_1+70.168761(b-y)m_1c_1 \nonumber \\
\end{eqnarray}

Figure 1 shows our estimates of $[{\rm Fe/H}]_{phot}$ arising from this calibration
plotted versus $[{\rm Fe/H}]_{spec}$ for all 1533 calibration measurements.

The CSR database has been compiled from a diverse array of metallicity
determinations, some of which have reported uncertainties, and some of
which do not. Furthermore, many of the 664 stars which survived our cuts
for inclusion into the calibration
are represented by multiple spectroscopic measurements. We circumvented
the non-uniform property of the CSR compilation by assuming that every
spectroscopic metallicity
measurement has an associated uncertainty of 0.05 dex. With
this assumption, each of the 1533 measurements contributes an equal weight to 
the calibration. We compute a reduced $\chi^{2}$ statistic of 6.68 for
the overall fit, which is superior to the reduced $\chi^{2}$ value of
9.33 which results from comparing $[{\rm Fe/H}]_{phot}$ obtained with
the SN calibrations with the spectroscopic metallicities.

In Figure 2, we plot the distribution of metallicity differences between
the photometric and spectroscopic determinations,
$[{\rm Fe/H}]_{phot}-[{\rm Fe/H}]_{spec}$, for both our calibration
and also for the SN calibration. Both distributions are well
represented by Gaussians. The new calibration is superior to SN
in that it has a narrower half-width at half maximum (0.10 dex
as opposed to 0.13 dex), and is centered more closely to zero
(-0.027 dex as opposed to -0.049 dex). Figure 3 shows a similar
diagram in which we plot only the results of measurements which
yielded a spectroscopic metallicity
$[{\rm Fe/H}]>0.0$. Here again, our calibration is
significantly superior to SN, which is not surprising, as very few
high metallicity stars were included in the SN calibration, which was
geared primarily toward studies of population II stars. 
Our calibration still tends to underestimate the metallicity of
the most metal-rich stars in comparison to spectroscopic determinations,
but the discrepancy has been significantly reduced. As a specific
example, the planet-bearing star HD 145675 (14 Her), which is the
highest metallicity star in our subset of the CSR catalog, with
$[{\rm Fe/H}]=0.5$ (Gonzalez, Wallerstein \& Saar 1999), receives a
metallicity estimate of [Fe/H]=0.47 from our calibration, and [Fe/H]=0.13
from SN.

Our metallicity calibration should provide a useful sieve for selecting 
candidate stars to be added to radial velocity surveys. We
note here a handful of class IV and V stars for which (1) our new calibration
finds $[{\rm Fe/H}]_{phot}>0.3$, which (2) do not appear in the Laughlin (2000)
catalog, and which (3) have no previously announced planet: HD 88176,
HD 112164, HD 15942, HD 144585.

In addition to providing spectroscopic metallicities, the 
CSR catalog also lists effective temperature estimates for each
of the cited entries. Following the same procedure and applying the
same cuts as for our our $[{\rm Fe/H}]_{phot}$ calibration, we have also derived
a $uvby-T_{eff}$ calibration,

\begin{eqnarray}
[{\rm T_{eff}}]_{phot} = &&4434.8364+29839.391(b-y)-13967.518m_1-3041.0698c_1 \nonumber \\
&-&98624.604(b-y)^2-8963.3232m_1^2+10620.269c_1^2+63289.783(b-y)m_1 \nonumber \\
&-&10099.516(b-y)c_1+16204.483m_1c_1+91256.827(b-y)^3-45093.846m_1^3 \nonumber \\
&-&10580.081c_1^3-122571.68(b-y)^2m_1+16993.421(b-y)^2c_1 \nonumber \\
&+&111551.98m_1^2(b-y)-43418.471m_1^2c_1-5856.0002c_1^2(b-y) \nonumber \\
&+&15492.020c_1^2m_1-127.04874(b-y)m_1c_1 \nonumber \\
\end{eqnarray}

If we assume that the average measurement uncertainty of the spectroscopic
$T_{eff}$ determinations is 50K, the reduced $\chi^{2}$ of this calibration
is 3.17, indicating that $uvby$ photometry is quite adequate to providing
effective temperature estimates for a given star. As before, we also
construct a distribution of differences ${T_{eff}}_{phot}-{T_{eff}}_{spec}$,
and construct a gaussian fit to the distribution.
The half-width at half maximum for this distribution is 80.4 K, whereas the
center lies at -11.5 K.

\section{Toward a Planeticity Calibration}

Given the demonstrated connection between the metallicity of a parent star and
the detectable presence of a planetary companion, and given the power of the 
Levenberg-Marquardt minimization technique in producing empirical calibrations,
one can ask a rather provacative question: Is it possible to construct a
``planeticity'' calibration in which the chance of finding a readily detectable
planet around a given star is estimated directly from uvby photometry?

Cumming, Butler and Marcy (1999), have published a detailed analysis of 11
years of precision radial velocity measurements for 73 nearby stars. Within
this list of stars, there are currently 10 announced planets. We assign a planeticity ${\cal P}=1.0$ to stars with planets, and ${\cal P}=0.0$ to stars without. We then
use the Levenberg-Marquardt method to produce a uvby-planeticity calibration in
exact analogy to producing the metallicity and effective temperature calibrations.
This exercise leads to the following empirical calibration:
\begin{eqnarray}
[{\cal P}_{phot}] = &&168.92038-1367.1865(b-y)+1215.0485m_1-625.29367c_1 \nonumber \\
&+&2450.1524(b-y)^2+251.91349m_1^2+686.49412c_1^2-3024.1263(b-y)m_1 \nonumber \\
&+&3776.9231(b-y)c_1-3394.6622m_1c_1-838.40082(b-y)^3-347.90246m_1^3 \nonumber \\
&-&238.02961c_1^3+698.73217(b-y)^2m_1-3804.6842(b-y)^2c_1 \nonumber \\
&+&802.37174m_1^2(b-y)-454.17907m_1^2c_1-2429.7350c_1^2(b-y) \nonumber \\
&+&2468.8122c_1^2m_1+4574.5147(b-y)m_1c_1 \nonumber \\
\end{eqnarray}

The results of this calibration are shown in Figure 4, and indicate that 
the planeticity calibration is not particularly convincing. Nevertheless,
it is instructive to test the relation on a larger set of stars.
We queried the Hipparcos catalog for stars lying within 50 pc of the sun,
retaining those for which (1) uvby photometry exists, (2) fall
in the range $0.8 M_{\odot} < M_{\star} < 1.5 M_{\odot}$ (using the
stellar evolution models of Allende Prieto  \& Lambert 1999), and which (3) 
were not among the 73 Lick Survey stars used to build the planeticity calibration given above.
These cuts left 2056 stars, of which 46 have reported planets. We expect
that the majority of the remaining 2010 are currently under radial
velocity surveillance, and that relatively few easily detected planets
will emerge from this aggregate. Using our calibration, we compute
an average planeticity of 0.271 for the 46 known planet-bearing stars,
in comparison to an average planeticity of 0.113 for the 2010 stars
without reported planets. 

We conclude that uvby photometry likely holds little planeticity
information beyond that provided by the well-recognized metallicity
proxy. We are very optimistic, however, regarding the possibility
of producing a planeticity calibration which is based on spectroscopically
determined differential metal abundances within a particular star.  Recent work, in particular by Gonzalez et al (2001), indicates that planet-bearing stars may have systematically unusual abundance ratios for $[{\rm Al/Fe}]$, $[{\rm Mg/Fe}]$, and $[{\rm Na/Fe}]$. Debra Fischer (2002, personal communication) has recently completed detailed spectroscopic abundance analyses of more than 500 stars which are currently maintained in the Keck and Lick planet search lists. In a future project, we are hoping to replace the uvby indices with spectroscopically determined differential abundances, and thus derive a more definitive planeticity calibration.

\subsection{Acknowledgements} 

We would like to thank Debra Fischer and Geoff Marcy for 
useful conversations. We would also like to thank Craig Markwardt
for advice regarding his IDL routines {\it mpfit} \& {\it mpfitfun}. This
work was supported by faculty research funds granted by the
University of California, Santa Cruz.

\newpage
\begin{center}
\large Figure Captions
\end{center}

\figcaption{
$uvby$ metallicities for planet-bearing stars
obtained with our photometric calibration
in comparison with spectroscopically determined metallicities
from the CSR 2001 compilation; solid line is a 1:1 correlation
\label{fig1}}

\figcaption{
Distribution of residuals between photometrically and spectroscopically
determined metallicities, with Gaussian fits to the distributions overplotted
in the same line style. Our new photometric calibration is shown as a 
solid line; the SN calibration is shown as a dashed line.
\label{fig2}}

\figcaption{
Distribution of residuals between photometrically and spectroscopically
determined metallicities, with Gaussian fits to the distributions overplotted.
Only measurements yielding a spectroscopic determination of 
${\rm [Fe/H]}>0.0$ are plotted.
Our new photometric calibration is shown as a
solid line; the SN calibration is shown as a dashed line.
\label{fig3}}

\figcaption{
Distribution of calculated values of planeticity for the 73 nearby Lick Survey stars.  Stars without planets are shown as a solid line; stars with confirmed planets are shown as a filled histogram.
\label{fig4}}

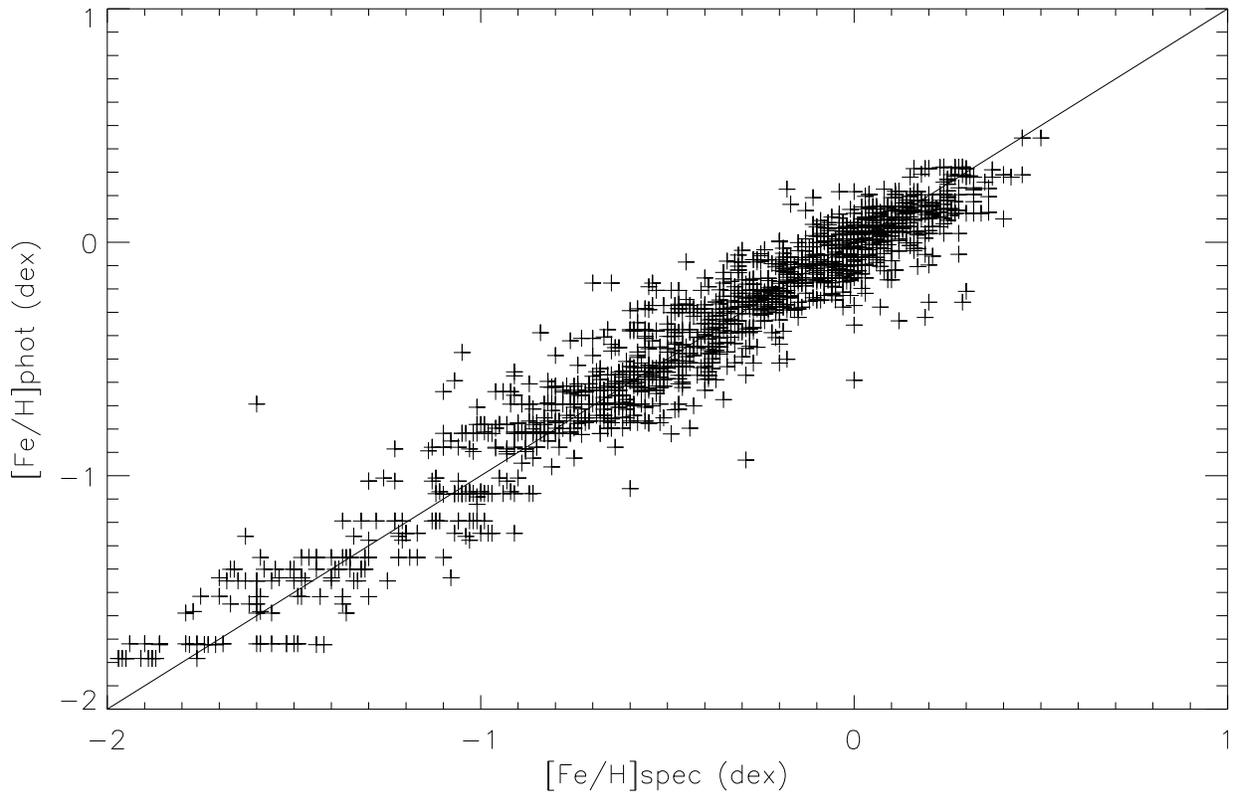
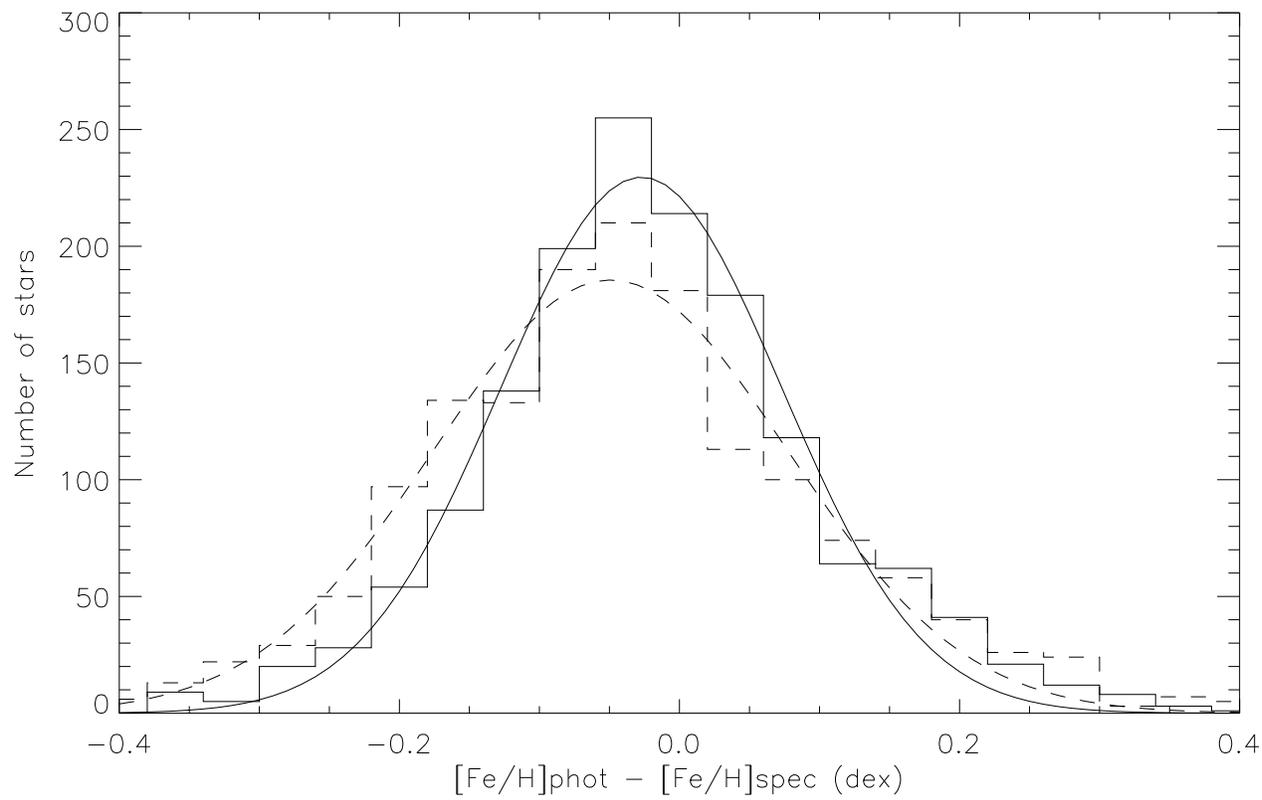
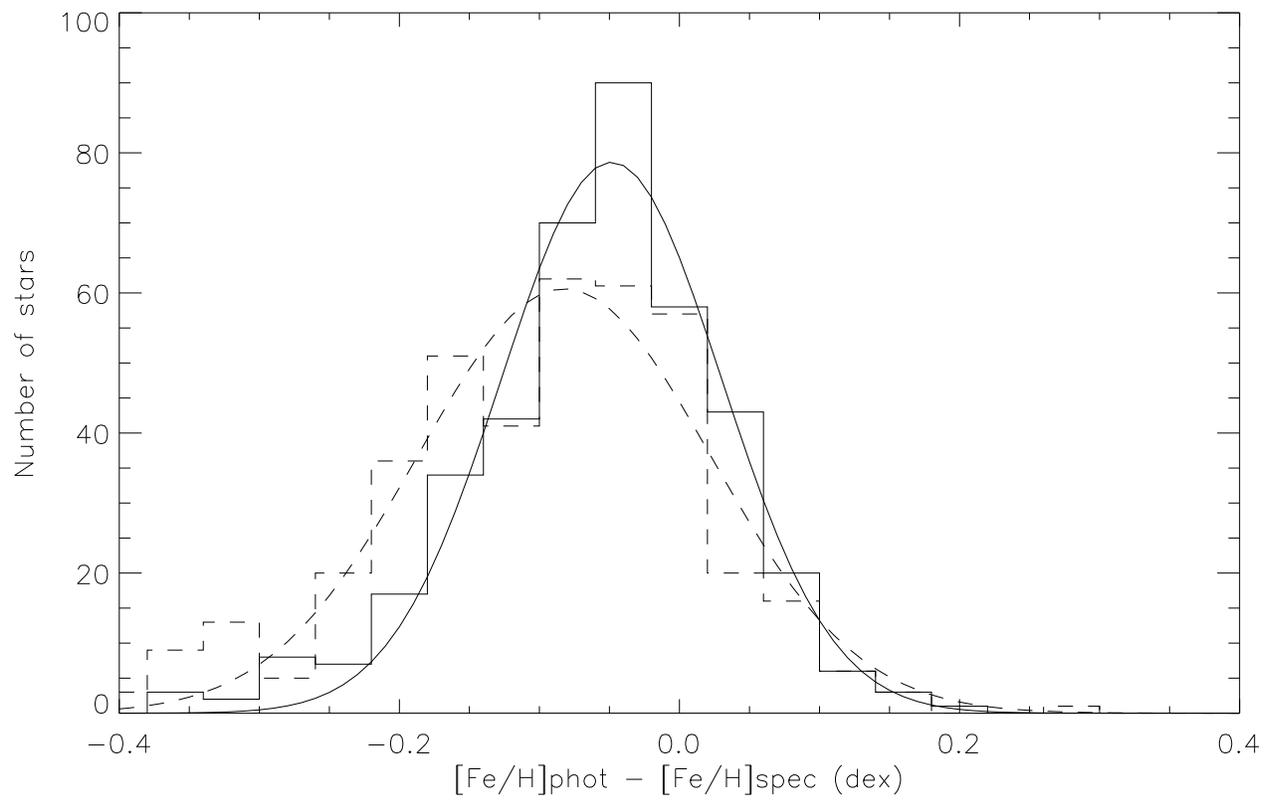
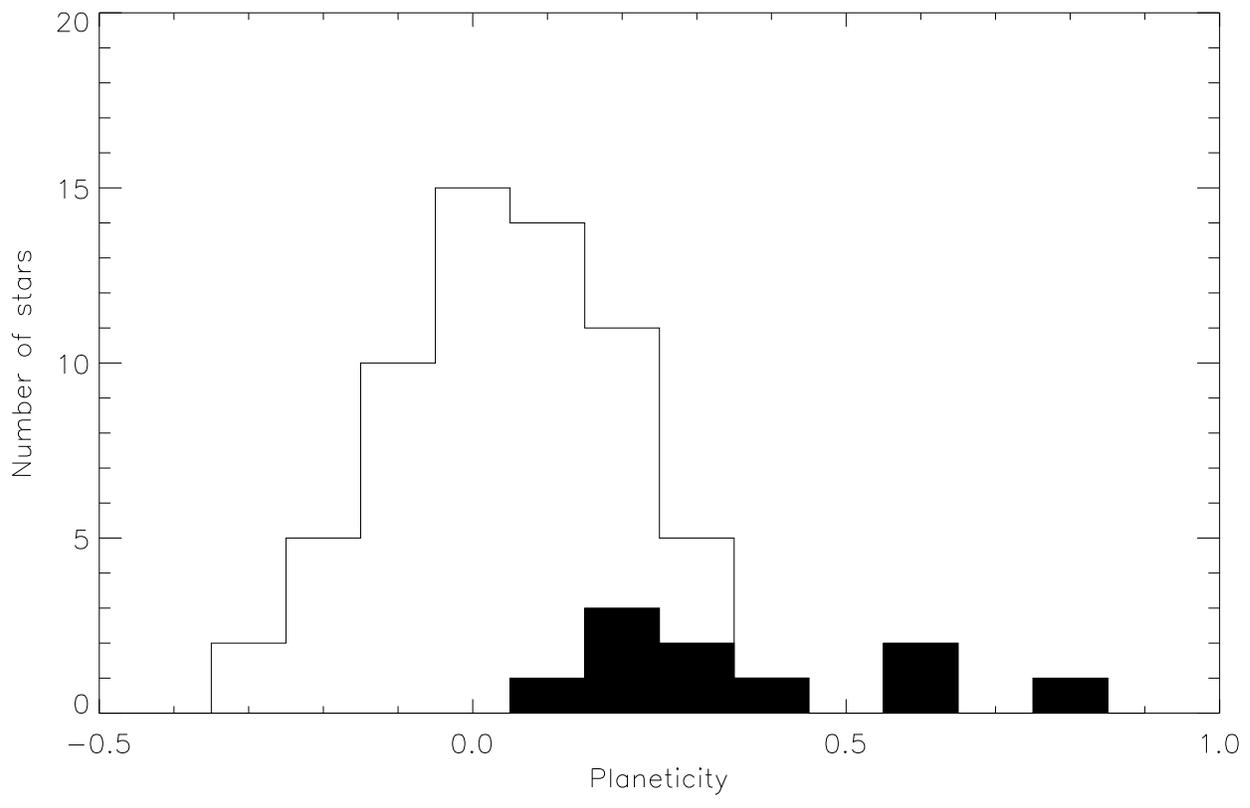
\end{document}